\documentclass[fleqn,twoside]{article}
\usepackage{espcrc2}
\readRCS
$Id: espcrc2.tex,v 1.2 2004/02/24 11:22:11 spepping Exp $
\usepackage{graphicx}
\newcommand{\AmS}{{\protect\the\textfont2
  A\kern-.1667em\lower.5ex\hbox{M}\kern-.125emS}}

\title{ %$\null$\\
Two-loop helicity amplitudes for fermion-fermion scattering}

\author{A. De Freitas \address{Deutsches Elektronen Synchrotron, DESY,
                       D-15738 Zeuthen, Germany}%
                      \thanks{Presenter. Alexander von Humboldt fellow.}
        Z. Bern \address{Department of Physics and Astronomy, UCLA,
                         Los Angeles, CA 90095-1547, USA}%
                \thanks{Research supported by the US Department of
                        Energy under grant DE-FG03-91ER40662.},
                             }

\def\bigglP{\biggl(}

\def\ksl{\not{\hbox{\kern-2.3pt $k$}}}

\def\e{\epsilon}

\def\Ord{{\cal O}}
\def\cm{{\cal M}}

\def\Nf{{N_{\! f}}}

\def\la{\langle}
\def\ra{\rangle}

\def\bom#1{{\mbox{\boldmath $#1$}}}

\def\tg{\tilde{g}}

\def\spa#1.#2{\left\langle#1\,#2\right\rangle}
\def\spb#1.#2{\left[#1\,#2\right]}
\def\lor#1.#2{\left(#1\,#2\right)}
\def\trc{{\rm Tr}}

\def\tS{{\tt S}}
\def\tT{{\tt T}}
\def\tU{{\tt U}}
\def\ibar{\bar{\imath}}

\begin{document}

\begin{abstract}
We discuss the calculation of two-loop helicity amplitudes for
quark-quark scattering in QCD and four-gluino scattering in ${\cal N}=1$
supersymmetric Yang-Mills theory. We study the dependence of the
results on different variants of dimensional regularization. In
particular, we consider the 't Hooft-Veltman and four-dimensional
helicity (FDH) schemes. We also discuss ambiguities in continuing the
Dirac algebra to $D$ dimensions. For the four-gluino case,
once the infrared divergent part of the amplitude is subtracted,
the finite remainder is free of these ambiguities.
\vspace{1pc}
\end{abstract}

\maketitle

The level of precision that will be achieved in experiments at the LHC
poses a challenge for theorists to match.  In principle, new physics
(or hints of it) may be found in small discrepancies between
theoretical predictions and experimental results and it is therefore
important that the theoretical predictions be robust.  As one
component of this, it would be helpful to know the cross section for
the production of hadronic jets through next-to-next-to-leading order
(NNLO) in the QCD coupling. There are a number of improvements to be
expected from such calculations, for example, reducing the
renormalization and factorization scale uncertainties in production
rates. It should also allow a better understanding of energy flows
within jets, as a jet may consist of up to to three partons at
NNLO. It also allows for better matching between parton-level and
hadron-level jet algorithms. Very importantly, it is also the first
order where honest assessments of theoretical uncertainties can be
made. See, for example, ref.~\cite{GloverReview} for further
discussions.

There has been great progress in the past few years in the calculation
of two-loop matrix elements, especially for $2 \to 2$ scattering
processes~\cite{BRY}-\cite{Gloverqqqq}.  This progress has been
possible thanks to new developments in loop integration
\cite{IBP}-\cite{IntegralsAGO} and in understanding the infrared
divergences of the theory \cite{Catani}-\cite{BDKtwoloopsplit}.  In
particular, in a series of papers the Durham group calculated the
interference of the two-loop amplitudes with the tree level ones,
summed over all external color and helicity states, for all $2 \to 2$
parton processes \cite{GOTYqqqq}-\cite{TwoLoopee3Jets}.

Two-loop amplitudes have also been calculated keeping full information
on the color and helicity states of the external
particles~\cite{TwoLoopee3JetsHel}-\cite{Gloverqqqq}.  This additional
information is not needed for the main phenomenological application,
namely, NNLO jet production in collisions of unpolarized
hadrons. However, experiments at RHIC involve the scattering of
polarized protons, for which the helicity amplitudes are directly
relevant.  Other advantages of having the amplitudes in a helicity
basis include the study of a number of formal properties of
scattering amplitudes, such as supersymmetry Ward identities
\cite{SWI}-\cite{TwoLoopSUSY}, collinear and high energy behavior
\cite{Neq4}-\cite{TwoloopBFKL}, and the link between color decomposed
QCD amplitudes, twistor space and topological string theories,
recently uncovered by Witten~\cite{Witten}.

We present here a summary of our calculation for quark-quark
scattering amplitudes in QCD, as well as four-gluino scattering in
${\cal N}=1$ super-Yang-Mills theory~\cite{qqqqpaper}. Our results for
the two-loop quark-quark helicity amplitudes agree with Glover's
results~\cite{Gloverqqqq} (after correction of minor errors).

The three QCD processes considered here are,
\begin{eqnarray}
&& \hskip-0.7cm  q(p_1,\lambda_1) + \bar{q}(p_2,\lambda_2)
\to \bar{Q}(p_3,\lambda_3) + Q(p_4,\lambda_4),
\label{qbQBlabel} \\
&& \hskip-0.7cm  q(p_1,\lambda_1) + \bar{Q}(p_2,\lambda_2)
\to q(p_3,\lambda_3) + \bar{Q}(p_4,\lambda_4),
\label{qBqBlabel} \\
&& \hskip-0.7cm  q(p_1,\lambda_1) + Q(p_2,\lambda_2)
\to q(p_3,\lambda_3) + Q(p_4,\lambda_4).
\label{qQqQlabel}
\end{eqnarray}
For the four-gluino case we consider
\begin{equation}
\tg(p_1,\lambda_1) + \tg(p_2,\lambda_2)
\to \tg(p_3,\lambda_3) + \tg(p_4,\lambda_4),
\label{gluinolabel}
\end{equation}
where we use a ``standard'' (not all outgoing) convention for the
external momentum $(p_i)$ and helicity labeling $(\lambda_i)$.

The amplitudes are calculated using dimensional regularization. We use
the following prescription when a trace of the Minkowski metric is
encountered,
\begin{equation}
\eta^{\mu}{}_{\mu} \equiv D_s \equiv 4 - 2 \e \, \delta_R  \,.
\label{EtaTrace}
\end{equation}
In this way, we have a continuous set of schemes labeled by
$\delta_R$. Setting $\delta_R=1$ corresponds to the 't Hooft-Veltman
scheme~\cite{HV}, while setting $\delta_R=0$ corresponds to the FDH
scheme~\cite{BKgggg,TwoLoopSUSY}.  The FDH scheme has improved supersymmetry
properties by virtue of fixing the number of internal gluon states to
two.

We evaluate the amplitudes using the spinor helicity
formalism~\cite{SpinorHelicity}.  In this formalism, the amplitudes
will be proportional to helicity-dependent phase-containing factors
written in terms of spinor inner products. These spinor inner products
are defined as $\la ij \ra = \la i^- | j^+ \ra$ and $[ij] = \la i^+ |
j^- \ra$, where $|i^\pm \ra$ are massless Weyl spinors of momentum
$p_i$, labeled with the sign of the helicity.

The $L$-loop amplitudes are color decomposed as,
\begin{equation}
 \cm^{(L)} = S \times
 \sum_{c=1}^3 \trc^{[c]} \times M^{(L),[c]} \,,
\label{RemoveColorPhase}
\end{equation}
where $S$ is the spinor factor mentioned above.
For example, for process (\ref{qbQBlabel}), with $\lambda_1=\lambda_4=+$ and
$\lambda_2=\lambda_3=-$, we have $S = i \la 31 \ra / \la 42 \ra$. The factors
for
other helicity configurations and for process (\ref{gluinolabel}) are similar.
The quantities $M^{(L),[c]}$ depend only on the Mandelstam variables
$s=(p_1+p_2)^2$,
$t=(p_1-p_4)^2$, and $u=(p_1-p_3)^2$.

For the quark process (\ref{qbQBlabel}) the color basis is
\begin{equation}
\trc^{[1]} = \delta^{i_4}_{~\ibar_1} \delta^{i_2}_{~\ibar_3}, \hskip0.5cm
\trc^{[2]} = \delta^{i_2}_{~\ibar_1} \delta^{i_4}_{~\ibar_3}.
\label{quarkbasis}
\end{equation}
The color bases for processes (\ref{qBqBlabel}) and (\ref{qQqQlabel})
are similar.  For the four-gluino case the color decomposition is the
same as for the four-gluon case, and is given in terms of traces of
color matrices (or products of them). Using the notation ${\rm
tr}(T^{a_i}T^{a_j}T^{a_k}T^{a_l}) = {\rm tr}_{ijkl}$ we can write
\begin{eqnarray}
&& \hskip-0.7cm \trc^{[1]}={\rm tr}_{1234}, \hskip0.4cm \trc^{[2]}={\rm
tr}_{1243},
\hskip0.4cm \trc^{[3]}={\rm tr}_{1423}
\nonumber \\
&& \hskip-0.7cm \trc^{[4]}={\rm tr}_{1324}, \hskip0.4cm \trc^{[5]}={\rm
tr}_{1342},
\hskip0.4cm \trc^{[6]}={\rm tr}_{1432},
\nonumber \\
&& \hskip-0.7cm \trc^{[7]}={\rm tr}_{12}{\rm tr}_{34},
   \hskip0.4cm \trc^{[8]}={\rm tr}_{13}{\rm tr}_{24},
\nonumber \\
&& \hskip-0.7cm \trc^{[9]}={\rm tr}_{14}{\rm tr}_{23}.
\end{eqnarray}

The two-loop Feynman diagrams were generated using {\tt QGRAF}
\cite{QGRAF}.  A {\tt MAPLE} program was then used to evaluate each
diagram. Some of the diagrams were also evaluated using {\tt FORM}
\cite{FORM} as a cross-check.

When the interference method is used, i.e., when one calculates the
two-loop amplitudes interfered with the tree-level ones, summed over
helicities and colors, one can use standard trace techniques to put
the integrand into a form containing only dot products of momenta.
Integral reduction algorithms then give the loop integrals in terms of
a minimal set of master integrals. In our case we want to keep full
information over helicity and color states. In order to put the
integrals into a form suitable for applying the general reduction
algorithms, we multiply and divide by appropriate spinor inner
products constructed from the external momenta. These spinor inner
products effectively play the role of the tree-level amplitudes in the
interference method, except that in this method the helicity
information is maintained when converting the spinor strings into
traces over $\gamma$ matrices.

An important way of checking the correctness of the calculation is by
comparing the infrared divergence of the renormalized amplitudes with
the ones predicted by Catani's formula for two-loop $n$-point
amplitudes \cite{Catani},
\begin{eqnarray}
&& | \cm_n^{(2)} \ra = {\bom I}^{(1)}
  \; | \cm_n^{(1)} \ra
%\nonumber \\ && \null
+ {\bom I}^{(2)} \;
         | \cm_n^{(0)} \ra \nonumber \\
&& \null \hskip 1.4 cm
 + |\cm_n^{(2){\rm fin}} \ra \,,
\label{TwoloopCatani}
\end{eqnarray}
where the ``ket'' notation $|\cm_n^{(L)} \ra$ indicates that the
$L$-loop amplitude is treated as a vector in color space. The
components of this vector are given by the $M_h^{(L),[c]}$ of
eq.~(\ref{RemoveColorPhase}).  The divergences of $\cm_n^{(1)}$ are
encoded in the color operator ${\bom I}^{(1)}$, while those of
$\cm_n^{(2)}$ also involve the scheme-dependent operator ${\bom
I}^{(2)}$. Catani's formula (\ref{TwoloopCatani}) not only allows us
to check the results, but also to organize them in terms of finite and
divergent parts.  Proofs of Catani's formulas have appeared in
refs.~\cite{StermanIR,BDKtwoloopsplit}.

For each process and each color basis we will have a different ${\bom
I}^{(1)}$ matrix.  For the basis (\ref{quarkbasis}) we have,
\begin{eqnarray}
&& \hskip-0.7cm {\bom I}^{(1)} =
{e^{-\epsilon \psi (1)} \over \Gamma (1-\epsilon )}
\left( {1\over\e^2} + {3\over 2\e} \right) \times
\label{I1} \\
&& \hskip-0.7cm
\left(
\begin{array}{c c}
2 C_F T - {1 \over N} (\tS - \tU)    &            \tT - \tU              \\
         \tS - \tU                   &  2 C_F \tS -{1 \over N}(\tT - \tU)
\end{array}
\right) \,,  \nonumber
\end{eqnarray}
where $N$ is the number of colors, $C_F = (N^2-1)/(2N)$ and
\[
  \tS = \left({\mu^2\over -s}\right)^\e \,, \hskip0.4cm
  \tT = \left({\mu^2\over -t}\right)^\e \,, \hskip0.4cm
  \tU = \left({\mu^2\over -u}\right)^\e \,. \hskip0.4cm
\]
The corresponding operator for $q\bar{Q} \to q\bar{Q}$ is obtained by
changing $\tS \to \tU $, $\tT \to \tS $ and $\tU \to \tT$ in
eq. (\ref{I1}). Similarly, the operator for $qQ \to qQ$ is
obtained by exchanging $\tS$ and $\tU$ in (\ref{I1}). For the
gluino case we get a nine-by-nine matrix, given in eq.~(2.18) of
ref. \cite{BDDgggg}.

The operator ${\bom I}^{(2)}$ is given by~\cite{Catani}
\begin{eqnarray}
&& {\bom I}^{(2)}
=  - \frac{1}{2} {\bom I}^{(1)}
\left( {\bom I}^{(1)} + {2 b_0 \over \e} \right) \nonumber \\
&& \null \hskip 1.1 cm
+ {e^{+\e \psi(1)} \Gamma(1-2\e) \over \Gamma(1-\e)}
\left( {b_0 \over \e} + K \right) {\bom I}^{(1)} \nonumber \\
&& \null \hskip 1.1 cm
+ {\bom H}^{(2)} \,,
\label{CataniGeneralI2}
\end{eqnarray}
where $b_0$ is the first coefficient of the QCD $\beta$-function, and
the coefficient $K$ depends on $\delta_R$ and
is given by~\cite{Catani,BDDgggg}
\begin{eqnarray}
&& K = \left[ \frac{67}{18} - \frac{\pi^2}{6}
    - \biggl( {1\over6} + {4\over9} \e \biggr) (1-\delta_R) \right] N
\nonumber \\ &&
\null \hskip 1. cm
- \frac{10}{18} \Nf \,,   \label{CataniK}
\end{eqnarray}
with $\Nf$ being the number of massless fundamental
representation quarks. ${\bom H}^{(2)}$ is a universal operator of order
$1/\e$. A general expression for ${\bom H}^{(2)}$ with an arbitrary number of
external
legs was recently presented in \cite{BDKtwoloopsplit}.

%\begin{center}
%\epsfxsize 2 in \epsfbox{closer2.ps}
%\end{center}

The two-loop remainders, as defined by eq. (\ref{TwoloopCatani}) will have
the following form
\begin{eqnarray}
M^{(2),[c]{\rm fin}} & \!\! = \!\! &
\biggl[ Q_1 -b_0^2 \left( \ln \left( {s \over \mu^2} \right) - i\pi \right)^2
\nonumber \\ && \phantom{\null \bigglP \null}
       -b_1 \left( \ln \left( {s \over \mu^2} \right) - i\pi \right)
       \biggr] M^{(0),[c]}
\nonumber \\ &&
\null + \biggl[ - 2b_0 \left( \ln \left( {s \over \mu^2} \right) - i\pi \right)
\nonumber \\ && \phantom{\null + \bigglP \null}
          +Q_2 \biggr] M^{(1),[c]{\rm fin}}
\nonumber \\ &&
\null + Q_3 M^{(1),[c],\e,\delta_R} +P^{[c]}\,,
\end{eqnarray}
where $Q_1$, $Q_2$ and $Q_3$ depend on the number of colors $N$,
number of flavors $\Nf$, and scheme label $\delta_R$ (in particular,
$Q_1=Q_2=Q_3=0$ when $\delta_R=1$). The parameter $\mu$ is the
renormalization scale. The $M^{(1),[c]{\rm fin}}$ are the one-loop
finite remainders, and the $M^{(1),[c],\e,\delta_R}$ are the
$\delta_R$-dependent parts of the $\Ord(\e^1)$ coefficients of the
one-loop amplitudes.  Finally, $P^{[c]}$ consists of powers of $N$ and
$\Nf$ multiplying functions (logarithms, polylogarithms, as well as
rational functions) of the Mandelstam variables.  The explicit form of
these functions is given in ref.~\cite{qqqqpaper}.

For the four-gluino case in ${\cal N}=1$ super-Yang-Mills theory, the two-loop
finite remainders
in the FDH scheme ($\delta_R=0$) are,
\begin{eqnarray}
M^{(2),[c]{\rm fin}} &\! \! = \! \!&
\biggl[ -\tilde{b}_0^2
\left(\ln \left( {s \over \mu^2} \right) - i\pi \right)^2
\nonumber \\ && \phantom{\null \bigglP \null}
       -\tilde{b}_1^2 \left( \ln \left( {s \over \mu^2} \right) - i\pi \right)
       \biggr] M^{(0),[c]}
\nonumber \\ &&
- 2\tilde{b}_0 \left(\ln \left( {s \over \mu^2} \right)
  - i\pi \right)  M^{(1),[c]{\rm fin}}
\nonumber \\ &&
+ N^2 A^{[c]} +B^{[c]} \,,
\label{SUSYfinite1}
\end{eqnarray}
for $c=1 \ldots 6$, and
\begin{eqnarray}
M^{(2),[c]{\rm fin}} &\!\!=\!\!&
- 2\tilde{b}_0
 \left( \ln \left( {s \over \mu^2} \right) - i\pi \right)  M^{(1),[c]{\rm fin}}
\nonumber \\ &&
\null + N G^{[c]} \,,
\label{SUSYfinite2}
\end{eqnarray}
for $c=7,8,9$. Here $\tilde{b}_0$ and $\tilde{b}_1$ are the first two
coefficients of the ${\cal N}=1$ super-Yang-Mills
$\beta$-function. $A^{[c]}$, $B^{[c]}$ and $G^{[c]}$ are functions of
$s$, $t$ and $u$.  Again the explicit form of all these function
is given in ref.~\cite{qqqqpaper}.

In previous papers~\cite{TwoLoopSUSY,BDDgggg,BDDqqgg}
we showed that the following supersymmetry Ward
identities are satisfied at two loops through $\Ord(\e^0)$
when one uses the FDH scheme:
\begin{eqnarray}
&& \hskip-0.6cm \cm^{\rm SUSY}(g_1^\pm,g_2^-,g_3^+,g_4^+)\ =\ 0,
\label{SUSYVanish} \\
&& \hskip-0.6cm \cm^{\rm SUSY}(\tg_1^+,\tg_2^-,g_3^+,g_4^+)\ =\ 0,
\label{SUSYVanishGluinos} \\
&& \hskip-0.6cm \cm^{\rm SUSY}(\tg_1^+,\tg_2^-,g_3^-,g_4^+)   \nonumber \\
&& \hskip .5 cm
\null =\
 { \spa2.3 \over \spa1.3 } \, \cm^{\rm SUSY}(g_1^+,g_2^-,g_3^-,g_4^+) \,.
\label{SUSYNonVanish}
\end{eqnarray}
With the four-gluino amplitudes, we may also check
the supersymmetry identity relating the four-gluon amplitude to the
four-gluino one,
\begin{eqnarray}
&& \hskip-0.6cm \cm^{\rm SUSY}(\tg_1^+,\tg_2^-,\tg_3^-,\tg_4^+)\ \nonumber \\
&& \hskip .5 cm
 =\ { \spa2.4 \over \spa1.3 } \, \cm^{\rm SUSY} (g_1^+,g_2^-,g_3^-,g_4^+).
\label{SUSYFourGluinos}
\end{eqnarray}
(Note that in ref.~\cite{BDDgggg} an all outgoing definition of
helicity is used for the four-gluon amplitudes, while we use the
`standard' one here where legs 1 and 2 are incoming.)  It turns out
that this last identity does not work immediately at two
loops~\cite{qqqqpaper}.  The problem is related to ambiguities arising
from continuing the Dirac algebra to $D$ dimensions.  At one-loop the
ambiguity is harmless because it affects only $\Ord(\e)$ terms.  From
the Catani formula a one-loop ambiguity at $\Ord(\e)$ leads to an
ambiguity in the $\Ord(1/\e)$ terms at two loops.  It is of course
possible to arrange for a prescription to fix the ambiguities to
restore the manifest supersymmetry Ward identities, but, in any case,
these two-loop ambiguities may all be absorbed into Catani's formula
for the divergent parts, leaving well defined finite parts.  As
expected, in the FDH scheme the finite remainders $A,B$ and $G$ in
eqs.~(\ref{SUSYfinite1}) and (\ref{SUSYfinite2}) satisfy supersymmetry
identities and agree with the corresponding functions given in
ref.~\cite{BDDgggg} for pure glue scattering.

The work of A.D.F. was supported by the Alexander von Humboldt Foundation.
We thank Lance Dixon for key contributions at early stages of this project.
We also thank Nigel Glover for communications regarding ref.~\cite{Gloverqqqq}.

%%%%%%%%%%%%%%%%%%%%%%%%%%%%%%%%%%%%%%%%%%%%%%%%

\end{document}